\documentclass[usegraphicx]{mn2e}
\usepackage{epsfig}
\usepackage{psfig}

\title[The brightness temperature problem in extreme IDV quasars: a model for PKS~0405-385]{The brightness temperature problem in extreme IDV quasars: a model for PKS~0405-385}
\author[R. J. Protheroe]{R. J. Protheroe\thanks{email: rprother@physics.adelaide.edu.au} \\
Department of Physics and Mathematical Physics, The University of Adelaide, Adelaide, SA 5005, Australia}

\begin{document}

\maketitle

\begin{abstract}
I re-examine the brightness temperature problem in PKS~0405-385
which is an extreme intra-day variable radio quasar with an
inferred brightness temperature of $\sim 5 \times 10^{14}$~K at
5~GHz, well above the Compton catastrophe limit of $\sim
10^{11}$~K reached when the synchrotron photon energy density
exceeds the energy density of the magnetic field.  If one takes
into account the uncertainty in the distance to the ionized
clouds responsible for interstellar scintillation causing rapid
intra-day variability in PKS~0405-385 it is possible that the
brightness temperature could be as low as $\sim 10^{13}$~K at
5~GHz, or even lower.  The radio spectrum can be fitted by
optically thin emission from mono-energetic electrons, or an
electron spectrum with a low-energy cut-off such that the
critical frequency of the lowest energy electrons is above the
radio frequencies of interest.  If one observes optically thin
emission along a long narrow emission region, the average energy
density in the emission region can be many orders of magnitude
lower than calculated from the observed intensity if one assumed
a spherical emission region.  I discuss the physical conditions
in the emission region and find that the Compton catastrophe can
then be avoided using a reasonable Doppler factor.  I also show
that MeV to 100 GeV gamma-ray emission at observable flux levels
should be expected from extreme intra-day variable sources such
as PKS~0405-385.
\end{abstract}


\section{Introduction}

Rapid variability in intra-day variable  (IDV) sources is a
long-standing problem as it implies apparent brightness
temperatures in the radio regime which may exceed 10$^{17}$ K, or
requires relativistic beaming with extremely high Doppler
factors, coherent radiation mechanisms, or special geometric
effects (Wagner \& Witzel 1995).  Such high brightness
temperatures would be well above the ``Compton catastrophe''
limit $T_B < 10^{11}$~K imposed by inverse-Compton scattering
(Kellermann \& Paulini-Toth 1969, Slysh 1992, Kardashev 2000)
when the photon energy density in the emission region exceeds the
energy density in the magnetic field.  See Krichbaum et al.\
(2002) and Kedziora-Chudczer et al.\ (2001) for recent reviews of
IDV sources.

The radio-loud quasar PKS~0405-385 is an extreme example of an
intra-day variable source with variations on timescales of
$t_{\rm IDV} \sim 0.1$~d (Kedziora-Chudczer et al.\ 1997).
Making the assumption that the emission region subtends solid
angle $\sim \pi(0.5ct_{IDV}D)^2/d_\theta^2$, where $D$ is the
Doppler factor and $d_\theta$ is the diameter distance to the
source, one can convert the observed 4.8~GHz flux to intensity,
and obtain a variability brightness temperature of $T_{\rm var}
\approx 10^{21}D^{-2}$~K.  In this source, however,
Kedziora-Chudczer et al.\ (1997) interpret the very short
variability time as due interstellar scintillation requiring the
angular diameter of the most compact component to be $\sim 6$
micro-arcsec, or smaller, corresponding to a solid angle
subtended by the IDV core $\Omega=\Omega_1$ where $\Omega_1
\approx 6.79 \times 10^{-22}$~sr.  The fraction of the total flux
they estimated to be associated with this compact IDV core is
$S_c \approx 0.15$, and the corresponding brightness temperature
at 4.8~GHz is $T_B \approx 5\times 10^{14}(\Omega_1/\Omega)$~K.
Although much lower than $T_{\rm var}$, to reconcile this
brightness temperature with the $\sim 10^{11}$~K limit would
appear to require a very large Doppler factor, $D \sim 10^3$, or
an even larger angular diameter.  To obtain a Doppler factor as
large as $10^3$ would require not only a jet Lorentz factor of
$\Gamma > D/2=500$, but also very close alignment of the jet axis
to our line of sight (within $\sim 1/\Gamma < 0.1^\circ$).  The
probability of such an alignment occurring by chance is then
$\sim 1/4\Gamma^2 < 10^{-6}$, in this case, which makes the very high
Doppler factor possibility unattractive.

The effective distance to the interstellar scintillation screen
is crucial in determining the angular size of the source, and
hence its brightness temperature.  The distance used by
Kedziora-Chudczer et al.\ (1997) and Walker (1998) is effectively
the scale-height above the Galactic plane of electron number
density squared, i.e.\ $z_2=\int_0^\infty z n_e^2(z)
dz/\int_0^\infty n_e^2(z) dz$, for which $z_2 \approx 500$~pc in
the model of Taylor and Cordes (1993) for the free electron
distribution in the Galaxy.  However, that model was developed
mainly for the consistent determination of pulsar distances from
dispersion measures, and is most accurate at low galactic
latitudes where the majority of pulsars are observed.  Taylor and
Cordes (1993) themselves warn that one should be aware of
uncertainties in their model associated with this.  The column
density, $\int_0^\infty n_e dz$, is fairly accurately determined, but
the scale height of electron number density, $z_1=\int_0^\infty z
n_e(z) dz/\int_0^\infty n_e(z) dz$, is less well determined, and $z_2$ is
even less accurately known.  For example, in the recent model of
Gomez et al.\ (2001) for the free electron distribution in the
Galaxy $z_2 \sim 300$~pc.  It is interesting to note that Beckert
et al.\ (2001) suggest typical distances to the scattering medium
of 200~pc, and that a scale height of about 100~pc seems to be
required to explain IDV in the case of 0917+624.

Krichbaum et al.\ (2002) mention the possibility of an extremely
clumpy ISM.  If this is the case, one could well question the use
of an effective screen distance as it could be that the
scintillation is due to an individual ionized cloud much nearer
to us than the average distance.  In fact, if the distribution is
highly peaked in the Galactic plane the most probable, rather
than average, distance is small.  The distance to the
ionized cloud responsible for the extreme scattering event in
0954+658 is estimated to be $\sim 150$~pc (Cimo 2002), while
Dennett-Thorpe \& de Bruyn (2000, 2002) suggest that the
scattering region for IDV in J1819+3845 may be located at about
20~pc, and possibly associated with the local bubble.  Hence, I
believe there could be considerable uncertainty in the distance
to the scintillating material responsible for the extreme IDV in
PKS~0405-385.  If, for example, the distance were smaller by a
factor of 5, this would translate into a factor of 25 reduction
in the brightness temperature.  A brightness temperature of $\sim
2 \times 10^{13}$~K at $\sim 5$~GHz for the IDV core in
PKS~0405-385 is still high, but I will show that it can be
achieved by standard electron synchrotron radiation using quite
moderate Doppler factors if one takes into account possible
geometries of the emission region.

In a recent paper (Protheroe 2002) I have explored the effect of
emission region geometry on flux variability, and on the
relationship between observed intensity and energy density for
various source geometries for the case of optically thin emission
and found that the average energy density in the source can be
much less than one would estimate simply from the observed
intensity.  Although the radio emission from IDV sources is
usually assumed to be optically thick, if this is not the case
then the above result may also have important implications for
IDV sources as the photon energy density responsible for causing
the brightness temperature limit may actually be a few orders of
magnitude lower than estimated from the intensity.  In that case,
lower Doppler factors would be required to avoid the Compton
catastrophe.  In this paper I shall explore the parameter space,
including emission region geometry, of models able to reproduce
the observed radio emission of the IDV core of PKS~0405-385, and
I shall model its spectral energy distribution (SED) from radio
to gamma-ray frequencies.

\section{Fitting the spectrum of PKS~0405-385}

\begin{figure}
\centerline{\psfig{file=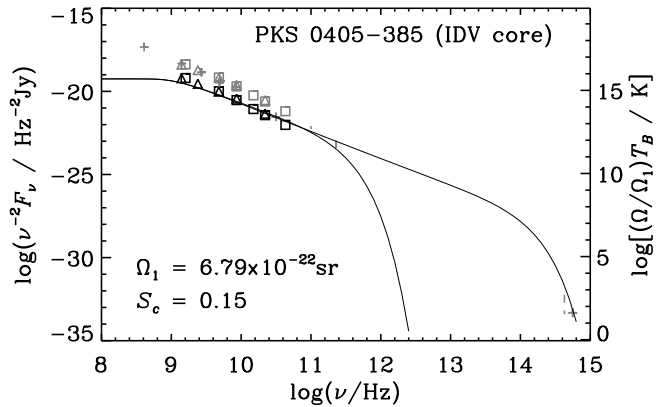,width=1.1\hsize}}
\caption{Observed fluxes of PKS~0405-385 divided by $\nu^2$ are
plotted as gray symbols and vertical lines.  Contemporaneous VLA
and ATCA data, multiplied by $S_c$, are also re-plotted as black
symbols and represent the flux from the IDV core.  The right-hand
axis shows the brightness temperature assuming the IDV core
subtends solid angle $\Omega_1$.  Data: pluses --
non-contemporaneous data obtained from the NED archive; triangles
and squares -- 1996 and 1998 VLA/ATCA data (D.\ Jauncey personal
communication 1999); vertical solid lines -- 1993-4 SEST data
(Tornikoski et al.\ 1996); dashed vertical lines at 230 GHz and
$4.3\times 10^{14}$~Hz -- (S.\ Wagner personal communication
1999).  The solid curves are fits for synchrotron radiation by
mono-energetic electrons with $\nu_1=10^{9.1}$~Hz and
$\nu_c=10^{11}$~Hz (left) or $\nu_c=10^{13.65}$~Hz (right).
\label{fig:TB}}
\end{figure}

In Fig.~\ref{fig:TB}, I plot the available flux measurements of
PKS~0405-385, divided by frequency squared, as gray symbols and
gray vertical lines.  Contemporaneous VLA and ATCA data (squares
and triangles in the 1.4~GHz--43~GHz range) are also re-plotted
as black symbols at the observed fluxes multiplied by $S_c$, the
fraction of the total flux assumed by Kedziora-Chudczer et al.\
(1997) to be associated with the IDV core.  The right axis shows
the brightness temperature of the IDV core inferred by
Kedziora-Chudczer et al.\ (1997) assuming the IDV core subtends
solid angle $\Omega_1$.  It is only the IDV core component of
these data that I am concerned with fitting.  Nevertheless, it is
interesting for comparison to include non-contemporaneous data at
other wavelengths whose origin may or may not be associated with
the same emission region.

Over the 1.4~GHz--43~GHz range, the intensity can be fitted well
by $I_\nu \propto \nu^{1/3}$ ($T_B \propto \nu^{-5/3}$).  Such a
spectrum would occur naturally if the emission were optically
thin and if this frequency range were well below the critical
frequency of the lowest energy electron.  Note that for an
XSelectron with Lorentz factor $\gamma$ the critical frequency is
\begin{eqnarray}
\nu_c = {3\gamma^2 e B_\perp \over 4\pi m_e c} = 4.19\times 10^6\gamma^2 B_\perp \mbox{ Hz}
\label{eq:nu_c}
\end{eqnarray}
where $e$ is the electron charge (statcoulombs), $B_\perp$ is the
component of magnetic field (gauss) perpendicular to the electron
velocity, and $m_e$ is the electron mass (grams).  Such an electron
spectrum could occur naturally if the electrons were produced as
secondaries of other particles, e.g.\ Bethe-Heitler
pair-production by relativistic protons, or if explosive
re-connection were responsible for electron acceleration.  For
the purposes of this paper, I shall adopt a mono-energetic
electron distribution.

For a mono-energetic isotropic electron distribution the synchrotron emission
coefficient is given by
\begin{eqnarray}
j_\nu = {P(\nu) \over 4\pi} = {\sqrt{3} e^3 B_\perp n_e \over 4\pi m_e c^2}F(x)
\end{eqnarray}
where $n_e$ is the
electron number density (cm$^{-3}$), $x=\nu/\nu_c$ and
$F(x)\equiv x\int_x^\infty K_{5/3}(\xi)d\xi$ (see e.g.\ Rybicki and Lightman 1979).
At low frequencies $\nu \ll \nu_c$
\begin{eqnarray}
F(x) = {4\pi \over \sqrt{3} \Gamma\!\left({1\over 3}\right)}\left({x \over 2}\right)^{1/3}, \;\;\;\;
j_\nu = {e^3 B_\perp n_e \over \Gamma\!\left({1\over 3}\right) m_e c^2}\left({x \over 2}\right)^{1/3}.
\label{eq:emiss_approx}
\end{eqnarray}

\begin{figure}
\centerline{\psfig{file=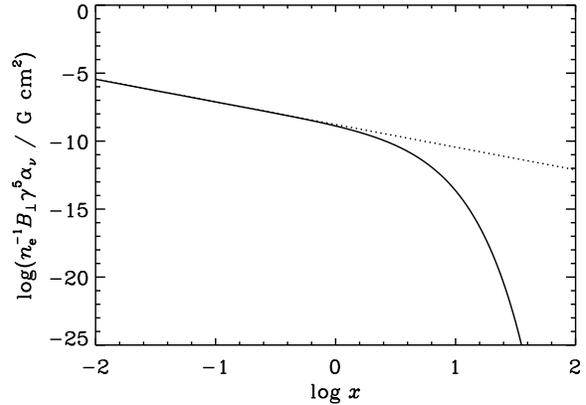,width=\hsize}}
 \caption{The synchrotron absorption coefficient vs.\ $x=\nu/\nu_c$ for 
mono-energetic electrons with Lorentz factor $\gamma$ and number
density $n_e$ (cm$^{-3}$) in a magnetic field with perpendicular
component $B_\perp$ (gauss).  The dotted line shows the
low-frequency limit $\alpha_\nu \approx 181.4 n_e
B_\perp^{2/3}\gamma^{-5/3}\nu^{-5/3}$.
\label{fig:synch_alpha}}
\end{figure}

The absorption coefficient for an isotropic electron distribution $N(E)$ is
\begin{eqnarray}
\alpha_\nu = {c^2 \over 8\pi h \nu^3}\int dE \; P(\nu,E)E^2\left[{N(E-h\nu) \over (E-h\nu)^2} -{N(E) \over E^2}\right].
\end{eqnarray}
For the mono-energetic electron distribution considered here $N(E)=n_e\delta(E-\gamma mc^2)$, and assuming $h\nu \ll E$, I find
\begin{eqnarray}
\alpha_\nu = - {4\pi e n_e \over 3^{3/2} B_\perp \gamma^5} {d \over dx}\left[{F(x)\over x}\right]
\end{eqnarray}
and this is plotted in Fig.~\ref{fig:synch_alpha}.
Note that at low frequencies
\begin{eqnarray}
\alpha_\nu & = & {32\pi^2 e n_e \over 27 B_\perp \gamma^5\Gamma\!\left({1\over 3}\right)2^{1/3}}x^{-5/3} \nonumber \\ & \approx & 181.4 n_e B_\perp^{2/3} \gamma^{-5/3} \nu^{-5/3} \mbox{ cm}^{-1}
\label{eq:alpha_approx}
\end{eqnarray}
and this is plotted as the dotted line in Fig.~\ref{fig:synch_alpha}.
From equations~(\ref{eq:emiss_approx}) and ~(\ref{eq:alpha_approx})
the source function at low frequencies is 
\begin{eqnarray}
S_\nu \equiv {j_\nu \over \alpha_\nu} \approx {3 \over 2} \gamma m_e \nu^2
\label{eq:source_fn}
\end{eqnarray}
which is identical to the Rayleigh-Jeans approximation to a black
body spectrum of temperature
\begin{eqnarray}
T={3 \gamma m_e c^2 \over 4 k}.
\label{eq:T_B}
\end{eqnarray}
Thus, in the optically-thick very-low frequency range the
brightness temperature is constant at $T_B \approx4.45 \times
10^9 \gamma$~K, and then there is a transition to $T_B \propto
\nu^{-5/3}$ at frequency $\nu_1$ where the optical depth
$\tau_\nu \equiv \int \alpha_\nu d \ell$ is unity, i.e.\
$\tau_\nu(\nu_1)\equiv 1$.  Hence, $T_B(\nu_1)$ can be used to
estimate the Lorentz factor of the electrons.

As noted earlier, the data appear to show $I_\nu \propto
\nu^{1/3}$ over the 1.4~GHz--43~GHz range corresponding to the
contemporaneous IDV observations.  At other frequencies it is not
known whether the observed emission is due to the same compact
component, or is due to a larger region, perhaps farther along
the jet.  However, data at other frequencies may still be used to
constrain the models.  Since the brightness temperature at low
frequencies is proportional to the electron Lorentz factor, the
brightness temperature problem is minimized by using the lowest
possible Lorentz factor.  I therefore take the highest frequency
$\nu_1$ which is just consistent with the 1.4~GHz data, and adopt
$\nu_1=10^{9.1}$~Hz.  Apart from the normalization, for which I
choose to take $F_1 \equiv F_\nu(\nu_1) \approx 8.9 \times
10^{-25}$~erg cm$^{-2}$s$^{-1}$Hz$^{-1}$ (after multiplying by
$S_c \approx 0.15$), the other parameter determining the fit is
the value of the critical frequency $\nu_c$.  The lowest critical
frequency $\nu_c$ which is consistent with the 43~GHz radio data
is $\nu_c \approx 10^{11}$~Hz, and the highest critical frequency
consistent with the R-band data is $\nu_c \approx 10^{13.65}$~Hz,
and I shall adopt these two values in my modeling.  The resulting
brightness temperature fits are shown by the solid curves in
Fig.~\ref{fig:TB}.  The challenge is now to find a combination of
source parameters, i.e.\ Doppler factor, jet-frame magnetic
field, electron number density and emission region geometry, that
gives the required intensity, and is physically possible.

If one views the emission region along the jet axis, one obtains
the largest Doppler boosting for any given jet Lorentz factor,
and I shall assume this to be the case for PKS~0405-385.  This
may well be justified as such extreme IDV radio quasars are
extremely rare.  Generally, the lower the energy density of
synchrotron radiation photons in the emission region, the lower
will be the Doppler factor required to avoid the Compton
catastrophe on synchrotron photons.  However, as I shall show,
very large Doppler factors can cause a Compton catastrophe on
cosmic microwave background radiation (CMBR) photons.  I shall
discuss the dependence of synchrotron photon energy density on
the geometry of the emission region in the next section.

\section{Synchrotron photon energy density and and emission region geometry}

In a separate paper (Protheroe 2002) I have discussed the
influence of emission region geometry on photon energy density in
the emission region for the optically thin case.  Here I shall
extend this work to include optical depth effects and use a
cylindrical emission region.  I shall consider the case of a
cylinder of length $\ell$ and radius $r$, with uniform emissivity
$j_\nu$ and absorption coefficient $\alpha_\nu$, and determine the
photon energy density $\langle U_\nu \rangle$ averaged over the
volume of the cylinder.

At some arbitrary point $\vec{r}$ within the cylinder the
intensity from direction defined by unit vector $\hat{e}$ is
\begin{eqnarray}
I_\nu(\vec{r},\hat{e}) = {j_\nu \over \alpha_\nu}\{1 -
\exp[-\alpha_\nu x(\vec{r},\hat{e})]\}
\end{eqnarray}
where $x(\vec{r},\hat{e})$ is the distance from $\vec{r}$ to
the boundary of the cylinder in direction $\hat{e}$.  The energy density at $\vec{r}$ is
\begin{eqnarray}
U_\nu(\vec{r}) = c^{-1}\oint I_\nu(\vec{r},\hat{e}) d\Omega.
\end{eqnarray}
Using a Monte Carlo method one can sample a large number, $N_d$,
of directions $\hat{e}_i, i=1,\dots,N_d$, distributed
isotropically, and then set
\begin{eqnarray}
U_\nu(\vec{r}) \approx c^{-1} \sum_{i=1}^{N_d} I_\nu(\vec{r},\hat{e}_i) (4\pi/N_d).
\end{eqnarray}

The energy density averaged over the volume $V=\pi r^2 \ell$ of the  
cylinder is then
\begin{eqnarray}
\langle U_\nu \rangle = V^{-1} \int U_\nu(\vec{r}) dV.
\end{eqnarray}
Using a Monte Carlo method one can sample a large number, $N_p$,
of points $\vec{r}_k, k=1,\dots,N_p$, distributed uniformly
throughout the volume of the cylinder, and then set
\begin{eqnarray}
\langle U_\nu \rangle = N_p^{-1} \sum_{k=1}^{N_p} U_\nu(\vec{r}_k).
\end{eqnarray}
Hence,
\begin{eqnarray}
\langle U_\nu \rangle = {4\pi j_\nu \over N_d N_p \, c \, \alpha_\nu} \sum_{k=1}^{N_p} \sum_{i=1}^{N_d}
\left\{ 1 -
\exp[-\alpha_\nu x(\vec{r}_k,\hat{e}_i)]\right\},
\label{eq:uphot_tau}
\end{eqnarray}
and in this way one can calculate $\langle U_\nu
\rangle$ for various $\ell/r$ values, and absorption coefficients
$\alpha_\nu$.  Of course for emission in the jet frame, all the 
variables in equation~(\ref{eq:uphot_tau}) would be jet-frame variables.

While the average energy density inside the cylinder is fixed for
any set of $j_\nu, \alpha_\nu, \ell/r$, and $r$, the intensity
observed when viewing the cylinder can depend strongly on viewing
direction.  In order to obtain the highest observed intensity one
would look in a direction such that the projected area of the
cylinder is smallest.  Making the axis of the cylinder coincident
with the jet axis, this would be achieved if one viewed the
emission region at $\theta'=\pi/2$ if the cylinder was short
($\ell'/r \ll 1$), and for $\theta'=0$ if the cylinder was long
($\ell'/r \gg 1$), where primed coordinates correspond to
jet-frame variables.  I shall consider the latter case, i.e.\
viewing the emission down the jet axis ($\theta'=0$).  The
jet-frame intensity is then simply given by $I'_{\nu'} =
(j'_{\nu'}/\alpha'_{\nu'})[1 - \exp(-\tau'_{\nu'})]$ where
$\tau'_{\nu'}=\alpha'_{\nu'} \ell'$.

\begin{figure}
\centerline{\psfig{file=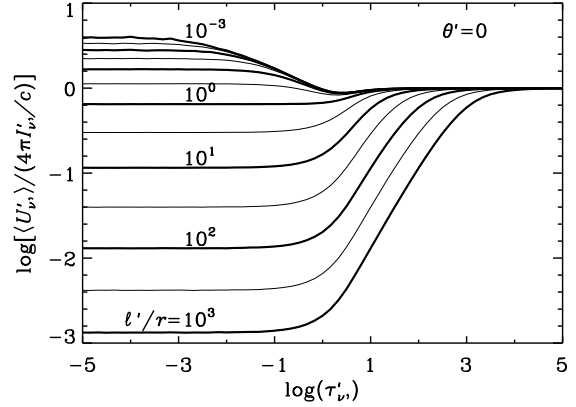,width=\hsize}}
 \caption{Average energy density inside the emission region
 divided by $4\pi/c$ times the observed intensity vs.\ optical
 depth for various $\ell'/r$.
\label{fig:uphot_tau}}
\end{figure}

In Fig.~\ref{fig:uphot_tau} I plot the ratio $\langle U'_\nu
\rangle /(4\pi I'_\nu/c)$ against $\tau'_{\nu'}$ for various values
of $\ell'/r$.  As can be seen, the effect is very important where
the emission is optically thin, e.g.\ for $\ell'/r=10^3$ the
average energy density is almost a factor $10^3$ lower than would
be expected from the observed intensity (note that the optical
depth is Lorentz invariant $\tau'_{\nu'}=\tau_\nu$).  In the next
section I shall consider various $\ell'/r$ values when exploring
the parameter space which could apply to the emission region for
IDV in PKS~0405-385.

\section{Model Parameters}

The fits shown by the solid curves in Fig.~\ref{fig:TB}
correspond to $F_1 \equiv F_\nu(\nu_1) \approx 8.9 \times
10^{-25}$~erg cm$^{-2}$s$^{-1}$Hz$^{-1}$, $\nu_1=10^{9.1}$~Hz, and
$\nu_c=10^{11}$~Hz or $10^{13.65}$~Hz.  In this section I shall
determine the combinations of physical parameters which could
give rise to these values, assuming a cylindrical emission
region geometry, an isotropic mono-energetic electron distribution
in the jet-frame, and observation along the jet axis.
The physical parameters describing the 
emission region are the perpendicular component of magnetic field
$B'_\perp$ as measured in the jet-frame, the Doppler factor $D$,
the ratio $\ell'/r$ of the emission region cylinder length to
its radius as measured in the jet-frame, and an
equipartition factor which gives the ratio of relativistic
electron energy density to magnetic energy density
\begin{eqnarray}
\eta=(\gamma' n'_e m_e c^2) \left({3 \over 2}{{B'}_\perp^2 \over 8\pi}\right)^{-1}
\label{eq:equilib}
\end{eqnarray}
where the (3/2) factor arises because $B'_\perp$ is the
perpendicular component of magnetic field, rather than its
magnitude, and $\gamma'$ is the jet-frame electron Lorentz factor.  

The solid angle subtended by the source is obtained from the
angular radius of the source, i.e.\ $\Omega=\pi\theta_{1/2}^2$,
and the radius of the cylindrical emission region is obtained
from the diameter distance and angular radius, i.e\
$r=d_\theta\theta_{1/2}$.  Note that for PKS~0405-385 at redshift
$z=1.285$ the diameter distance is $d_\theta=5.68\times
10^{27}$~cm for a $\Lambda$CDM model with $\Omega_m=0.2,
\Omega_\Lambda=0.8$ and $H_0=70$~km s$^{-1}$ Mpc$^{-1}$ (here
$\Omega$ specifies the fraction of the cosmological closure
density, whilst elsewhere in this paper $\Omega$ is used for solid
angles).

Noting that $I_\nu=F_\nu/\Omega$, and that in the optically thick
region of the spectrum $I_\nu=j_\nu/\alpha_\nu$, from equation~(\ref{eq:source_fn})
one may obtain the Lorentz factor the electrons would have if the
emission took place in the observer frame, $\gamma_1$, and hence find
the jet frame Lorentz factor
\begin{eqnarray}
\gamma' = \gamma_1/D = {2 \over 3} {F_1 \over D \pi\theta_{1/2}^2 m_e \nu_1^2} .
\label{eq:gamma_prime}
\end{eqnarray}
Next, using equation~(\ref{eq:alpha_approx}) and that, by definition, the optical
depth at jet-frame frequency $\nu_1'=\nu_1/D$ must be unity
\begin{eqnarray}
 181.4 && \hspace*{-2em}(\ell'/r) (\theta_{1/2}d_\theta) n'_e {B'}_\perp^{2/3} \times \nonumber \\ && \hspace*{0em}  \times \left( {2 F_1 \over 3 D \pi\theta_{1/2}^2 m_e \nu_1^2}\right)^{-5/3} \left( {\nu_1 \over D}\right)^{-5/3} = 1.
\end{eqnarray}
Substituting for $n_e'$ from equation~(\ref{eq:equilib}) with $\gamma'$ from equation~(\ref{eq:gamma_prime}) and solving for $B'_\perp$ I obtain
\begin{eqnarray}
B'_\perp \! &=& \! 4.98 \times 10^{23} D^{-13/8}  \nu_1^{-11/8} \theta_{1/2}^{-19/8}  \times  \nonumber \\ && \hspace*{6em}  \times F_1 d_\theta^{-3/8} \left(\eta {\ell'\over r}\right)^{-3/8} \mbox{~~G}.
\label{eq:b_from_nu1}
\end{eqnarray}
Then, using equation~(\ref{eq:nu_c}) for the  jet-frame critical frequency, $\nu_c'=\nu_c/D$, I obtain
\begin{eqnarray}
B'_\perp \! &=& \! 4.38 \times 10^{-60}   D \nu_1^{4} \nu_c \theta_{1/2}^{4} F_1^{-2}  \mbox{~~G}.
\label{eq:b_from_nuc}
\end{eqnarray}
Finally, solving simultaneous equations~(\ref{eq:b_from_nu1}) and (\ref{eq:b_from_nuc}) I obtain
\begin{eqnarray}
D \! &=& \! 4.37\!\times\! 10^{31} \nu_c^{-8/21} \nu_1^{-43/21}  \times  \nonumber \\ && \hspace*{4em}
 \times \theta_{1/2}^{-17/7} F_1^{8/7} d_\theta^{-1/7} \left(\eta
{\ell'\over r}\right)^{-1/7}
\label{eq:d_model}
\end{eqnarray}
\begin{eqnarray}
B'_\perp \! &=& \! 1.91\!\times\! 10^{-28} \nu_c^{13/21} \nu_1^{41/21} \theta_{1/2}^{11/7}  \times  \nonumber \\ && \hspace*{4em} \times 
F_1^{-6/7} d_\theta^{-1/7} \left(\eta
{\ell'\over r}\right)^{-1/7} \mbox{~~G.}
\label{eq:b_model}
\end{eqnarray}

\begin{figure}
\centerline{\psfig{file=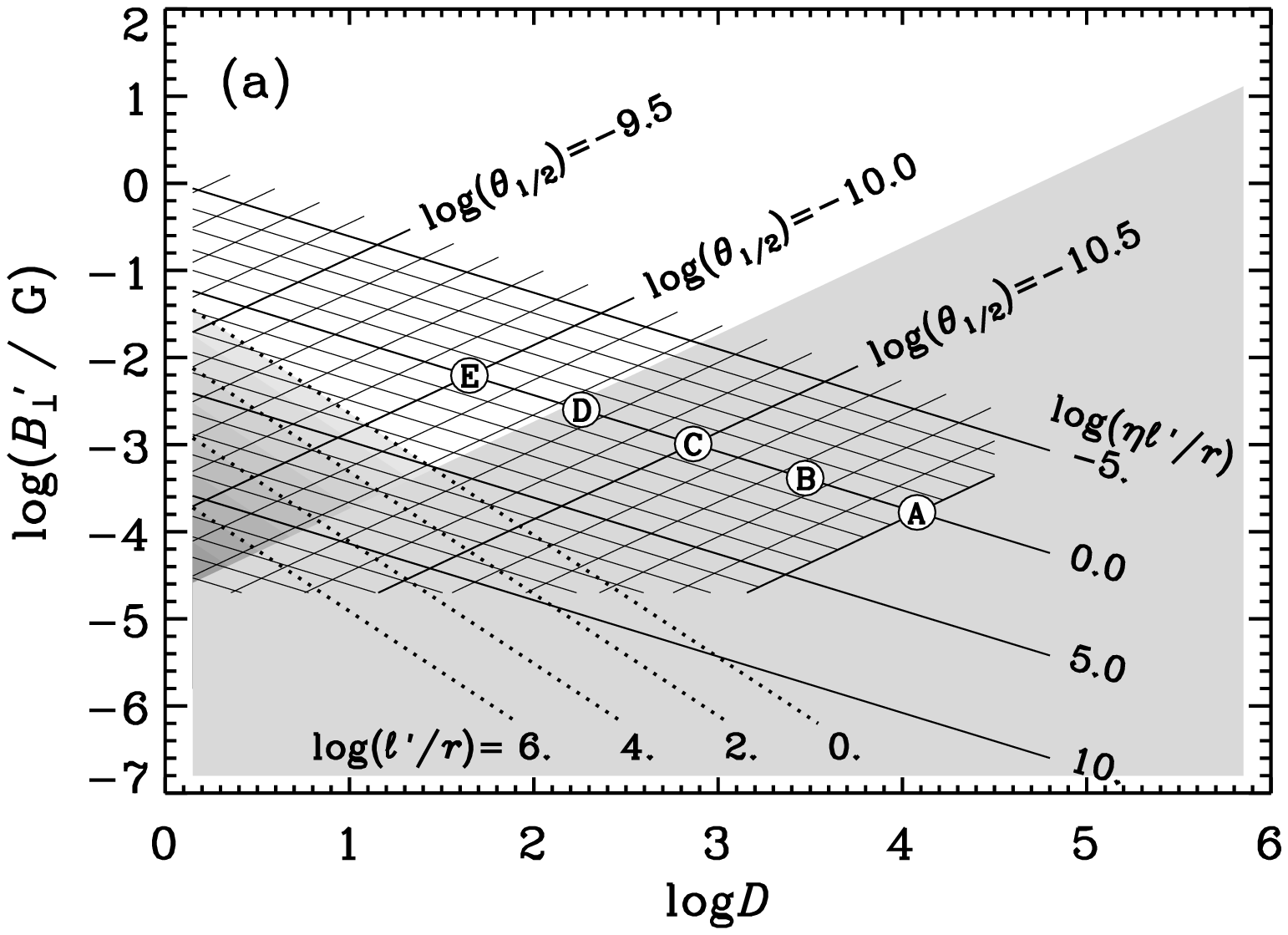, width=1.2\hsize}}
\centerline{\psfig{file=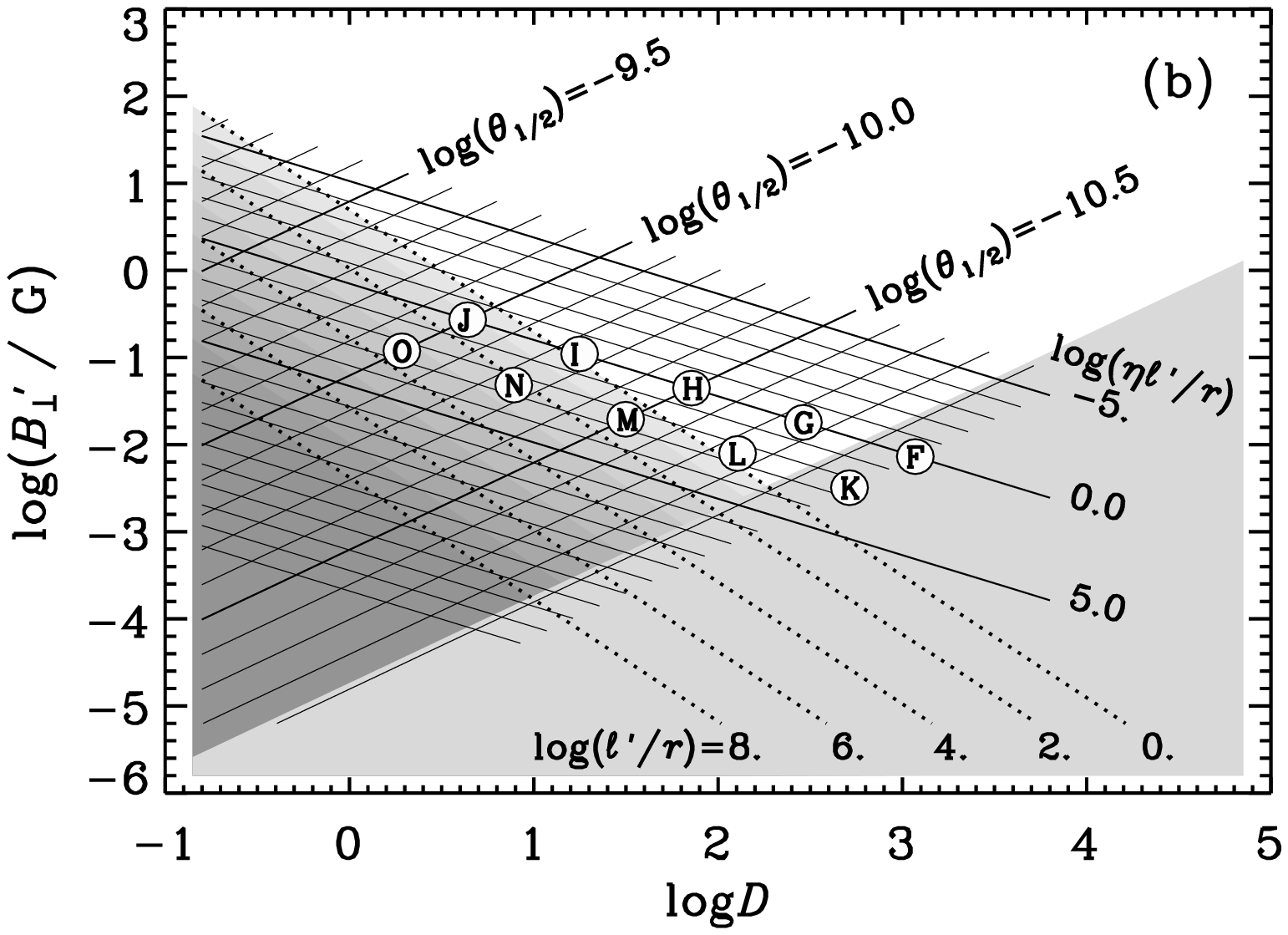, width=1.2\hsize}}
\caption{Magnetic field--Doppler factor parameter space for
models fitting the radio intensity of the IDV core of
PKS~0405-385 with $\nu_1=10^{9.1}$~Hz and (a) $\nu_c=10^{11}$~Hz
or (b) $\nu_c=10^{13.65}$~Hz, i.e.\ giving the left and right
solid curves in Fig.~1.  The two sets of solid curves labeled by
$\log(\eta\ell'/r)$ and $\log(\theta_{1/2})$ are given by
equations~(\protect\ref{eq:d_model}) and
(\protect\ref{eq:b_model}) plotted using $\theta_{1/2}$ and
$(\eta\ell'/r)$, respectively, as the parameter, and are used to
obtain the values of $\theta_{1/2}$ and $(\eta\ell'/r)$ for any
combination of $B_\perp$ and $D$ that are necessary to give the
required radio spectrum.  The shaded region on the right shows
where the energy losses are dominated by IC on the CMBR and
corresponds to equation~(\protect\ref{eq:b_min_ic}), and the
shaded regions on the left show where the energy losses are
dominated by IC on synchrotron photons and correspond to
equation~(\protect\ref{eq:b_min_ssc}) for various values of
$\log(\ell'/r)$ indicated.  Fifteen models, i.e.\ combinations of
$B_\perp$, $D$, $\theta_{1/2}$ and $(\eta\ell'/r)$, which will be
discussed later are labeled A to O.}
\label{fig:parameters}
\end{figure}

Fig.~\ref{fig:parameters} shows the magnetic field--Doppler
factor parameter space for models fitting the radio intensity of
the IDV core of PKS~0405-385.  Equations~(\ref{eq:d_model}) and
(\ref{eq:b_model}) define models which will give synchrotron
spectra of mono-energetic electrons determined by $F_1$, $\nu_1$
and $\nu_c$, and these models are represented as solid lines in
Fig.~\ref{fig:parameters} corresponding to either a fixed value
of $(\eta\ell'/r)$ while varying $\theta_{1/2}$ as a parameter in
these equations, or a fixed value of $\theta_{1/2}$ while varying
$(\eta\ell'/r)$.  All other variables are fixed at the values
appropriate to PKS~0405-385, with Fig.~\ref{fig:parameters}(a)
being for $\nu_c=10^{11}$~Hz and Fig.~\ref{fig:parameters}(b)
being for $\nu_c=10^{13.65}$~Hz ($\nu_1=10^{9.1}$~Hz in both
cases).  

If one allows the relativistic particle energy density to exceed
the magnetic energy density, then very low Doppler factors are
possible. However, if the equipartition factor is greater than
unity, this would be unstable.  It is therefore probably
unrealistic for the relativistic particles to be too far from
equipartition with the magnetic field, and so one should perhaps
take more seriously  results corresponding to $\eta=1$, which I shall
assume in what follows.
Nevertheless, IDV is truly a time-dependent problem, and
episodes with $\eta >1$ are not ruled out completely.

\subsection{Avoiding Compton catastrophes}

If electrons are injected with jet-frame Lorentz factor $\gamma'$
then the presence of dense radiation fields provides target
photons for inverse Compton scattering, and the generation of
components in the SED at X-ray and gamma-ray frequencies which
may or may not exceed observed X-ray flux and gamma-ray limit.  If
the electrons are accelerated in a quasi-continuous process such
as diffusive shock acceleration by non-relativistic or
mildly-relativistic shocks, then the presence of dense radiation
fields may lead to excessive energy losses and prevent the
electrons reaching the Lorentz factor required to fit the
observed radio spectrum.  In either case, these effects may occur
wherever the photon energy density in the emission region becomes
comparable to or exceeds the energy density in the magnetic field.

In addition to the synchrotron emission itself, I shall assume
that emission region is sufficiently far along the jet that the
only other non-negligible field for inverse-Compton scattering is
the cosmic microwave background radiation (CMBR).  If the
emission region were close to the central engine then accretion
disk radiation, broad line cloud emission, and torus infrared
emission could also play a role (Donea \& Protheroe 2002).

The average jet-frame energy density of the synchrotron emission
inside the emission region can be estimated for given $\ell'/r$
values from the observed intensity $I_\nu$ corresponding to
$\nu^2F_\nu$ plotted as the solid curves in Fig.~\ref{fig:TB}.
The energy density in the jet frame is given by
\begin{eqnarray}
\langle U'_{\rm syn}(\ell'/r,\theta_{1/2}) \rangle = D^{-4} \langle U_{\rm
syn}(\ell'/r,\theta_{1/2}) \rangle
\end{eqnarray}
where 
\begin{eqnarray}
&& \hspace*{-6em} \langle U_{\rm syn}(\ell'/r,\theta_{1/2})
\rangle \nonumber \\ \hspace*{3em} &=& {4\pi \over c}\int
M[\ell'/r,\tau_\nu(\nu)] I_\nu(\nu,\theta_{1/2}) d\nu,
\end{eqnarray}
$\tau_\nu(\nu) \approx (\nu_1/\nu)^{5/3}$ from
Fig.~\ref{fig:synch_alpha}, and $M[\ell'/r,\tau_\nu(\nu)]=\langle
U'_{\nu'}\rangle /(4\pi I'_{\nu'}/c)$ given in
Fig.~\ref{fig:uphot_tau}.  The main contribution to the energy
density for the spectrum fitted to the radio observations of
the IDV core in PKS~0405-385 comes from frequencies near to
$\nu_c$ since this is where the SED, $\nu
F_\nu$, peaks.  This spectrum is optically thin near $\nu_c$ and
so I may make the approximation
\begin{eqnarray}
&& \hspace*{-6em} \langle U_{\rm syn}(\ell'/r,\theta_{1/2}) \rangle \nonumber \\ 
\hspace*{3em} & \approx & M(\ell'/r,\tau_\nu \ll 1){4\pi \over c}\int I_\nu(\nu,\theta_{1/2}) d\nu,  \nonumber \\ 
\hspace*{3em} & \approx & M(\ell'/r,\tau_\nu \ll 1) g(\nu_1,\nu_c) F_1 \theta_{1/2}^{-2},
\end{eqnarray}
where
\begin{eqnarray}
g(\nu_1,\nu_c) = {4 \over c}\int {F_\nu(\nu) \over F_\nu(\nu_1)}d\nu.
\end{eqnarray}
For the spectra shown by the solid curves in Fig.~\ref{fig:TB}, $M(1,\tau_\nu \!\ll\! 1) g(10^{9.1}$Hz$,10^{11}$Hz$)=45.14$~cm$^{-1}$sr~Hz and $M(1,\tau_\nu \! \ll \!1) g(10^{9.1}$Hz$,10^{13.65}$Hz$)=1.47 \times 10^5$~cm$^{-1}$sr~Hz.

In the rest frame of the host galaxy the CMBR would have had
temperature $(1+z)T_0$, where $T_0 = 2.735$~K, and would
have been isotropic.  I shall distinguish between the Doppler
factor for boosting from the jet frame to the host galaxy
frame, $\delta$, and the Doppler factor for boosting from the
jet frame to the observer frame, $D=\delta/(1+z)$.  In the jet
frame, the black body temperature would depend on the angle $\psi'$
with respect to the jet axis at which an observer in the jet frame
looks
\begin{eqnarray}
T'(\psi')={(1+z)T_0 \over \Gamma(1 - \beta_j \cos\psi')}
\end{eqnarray}
where $\beta_j c$ is the jet velocity as measured in the host galaxy
frame, $\Gamma=(1-\beta_j^2)^{-1/2}$ is its Lorentz factor, and is given by
$\Gamma=(\delta^2+1)/(2\delta)$ for the case of observation of the AGN jet at angle $\theta=0$ to its axis.  The energy
density of the CMBR in the jet frame is then
\begin{eqnarray}
U'_{\rm CMBR} &=& {(1+z)^4a T_0^4 \over 4
\pi}\oint{d\Omega' \over [\Gamma(1 - \beta_j \cos\theta')]^4} \nonumber \\  &=& {4\over
3}\Gamma^2(1+z)^4a T_0^4 .
\end{eqnarray}
where $a=4\sigma/c$ and $\sigma$ is the Stefan-Boltzmann
constant.  Note that whereas the jet frame energy density of
synchrotron radiation decreases with assumed Doppler factor, the
energy density of the CMBR increases with assumed Doppler factor.
This means that the Compton catastrophe can not be avoided by
using an arbitrarily large Doppler factor.  In fact, very high
Doppler factors will result in a Compton catastrophe due to
inverse-Compton scattering on the CMBR.  This has important
consequences which I shall address next, and in subsequent
sections.

For efficient synchrotron radiation to take place the
energy density in the magnetic field should exceed the energy
density in the radiation field
\begin{eqnarray}
{3\over 2}{{B'}_\perp^2\over (8\pi)} > \left[D^{-4} \langle U_{\rm
syn}(\ell'/r) \rangle + {4\over
3}\Gamma^2(1+z)^4a T_0^4\right].
\label{eq:b_min} 
\end{eqnarray}
Since the synchrotron photon energy density is large in the
regime of low Doppler factors, while the CMBR energy density is
large in the regime of high Doppler factors, I shall obtain
separately the minimum magnetic field required in each case.  For
PKS~0405-385 being viewed along the jet axis, and using the
approximation $\Gamma \approx (1+z)D/2$, valid for $\Gamma \gg
1$, the jet-frame energy density of the magnetic field is less
than that of the CMBR when
\begin{eqnarray}
B'_\perp < (4/3)\sqrt{a\pi} (1+z)^3 T_0^2 D
\label{eq:b_min_ic} 
\end{eqnarray}
and this is independent of all model parameters except $D$.
This case is shown by the shaded area on the right in 
Figs.~\ref{fig:parameters}(a)--(b).

In the case of synchrotron photons as targets for inverse Compton scattering,
i.e.\ the synchrotron self-Compton (SSC) process, inverse Compton   losses dominate when
\begin{eqnarray}
{B'}_\perp &<& 4(\pi/3)^{1/2}D^{-2}\langle U_{\rm syn}(\ell'/r,\theta_{1/2})  \rangle^{1/2}.
\label{eq:b_min_ssc}
\end{eqnarray}
Because in this case the photon energy density depends on $\theta_{1/2}$
and $\ell'/r$, in addition to $D$, the boundary between models
($D,B'_\perp$) is parametrized by substituting $D$ from
equation~(\ref{eq:b_from_nuc}) into equation~(\ref{eq:b_min_ssc}) and
solving for $B'_\perp$ as a function of the parameters, and then
using equation~(\ref{eq:b_from_nuc}) to obtain $D$ as a function of
the parameters.  
The resulting minimum value of $B'_\perp$ is
plotted against $D$ in Figs.~\ref{fig:parameters}(a)--(b) for
various values of $\ell'/r$ as the dotted lines (labeled by
$\ell'/r$) bounding the shaded areas on the left.

The gyroradius, $r'_g =\gamma'c/(B'_\perp\omega_B)$ must be much
smaller than the radius $r$ of the cylinder and so this provides,
in principle, an additional lower limit to $B'_\perp$.  However,
this limit is well below the lower limit from
equation~(\ref{eq:b_min}) already plotted, and therefore does not
further constrain the models.

Fifteen potential models which will be discussed later, i.e.\
combinations of $B_\perp$, $D$, $\theta_{1/2}$ and
$(\eta\ell'/r)$, are labeled as A to O in
Fig.~\ref{fig:parameters}.  Before proceeding, we should check if
any of these models is ruled out by being optically thick to
Thomson scattering.  The Thomson optical depth is $\tau_T = \ell'
n_e' \sigma_T$ where $\sigma_T$ is the Thomson cross section.
Taking $n_e'$ from equation~(\ref{eq:equilib}) with $\gamma'$
from equation~(\ref{eq:gamma_prime}) I obtain
\begin{eqnarray}
\tau_T = 2.11 \times 10^{24} {B'}_\perp^2 D  \theta_{1/2}^3 \left(\eta
{\ell'\over r}\right).
\end{eqnarray}
For the fifteen potential models the Thomson optical depth  ranges from
$\tau_T=7.04 \times 10^{-13}$ (Model A) to $\tau_T=1.83 \times
10^{-5}$ (Model O), and so Thomson scattering can be neglected in
this case.

\section{Calculating the Spectral Energy Distribution}

I shall consider the following three emission processes:
synchrotron radiation (syn), inverse Compton scattering of synchrotron
photons (SSC), and inverse Compton scattering of CMBR photons (ICM).
Assuming the inverse Compton scattering is in the Thomson regime,
the rate of energy loss by either process is
\begin{eqnarray}
{dE' \over dt'} = - {4 \over 3} \sigma_T c {\gamma'}^2 U'
\label{eq:dedt}
\end{eqnarray}
where $U'=U'_B$ for synchrotron losses, $U'=\langle U'_{\rm syn}
\rangle$ for SSC losses and $U'=U'_{\rm CMBR}$ for
inverse-Compton losses on the CMBR.  The fraction of emitted
radiation by each process is then
\begin{eqnarray}
f_{\rm syn} &=& U'_B/(U'_B+\langle U'_{\rm syn} \rangle+U'_{\rm CMBR})\\ 
f_{\rm SSC} &=& \langle U'_{\rm syn} \rangle/(U'_B+\langle U'_{\rm syn} \rangle+U'_{\rm CMBR})\\
f_{\rm ICM} &=& U'_{\rm CMBR}/(U'_B+\langle U'_{\rm syn} \rangle+U'_{\rm CMBR}).
\end{eqnarray}
This does not mean, however, that the observed energy flux for
the inverse-Compton scattered CMBR is in this ratio to the other
two components because, whereas the synchrotron emission and SSC
emission is isotropic in the jet frame, the inverse-Compton
scattering of the CMBR is not because of the anisotropy of the
CMBR in this frame.  One may well expect strong peaks in the SED
due to inverse-Compton scattering for models close to or within
the shaded areas in Figs.~\ref{fig:parameters}(a)--(b).  Slysh
(1992) also noted that gamma-ray emission may occur in IDV
sources.

\subsection{Synchrotron radiation}

The intensity of synchrotron emission in direction $\theta=0$ in
the observer frame is simply given by
\begin{eqnarray}
I_{\nu}^{\rm syn}(\nu,\theta=0) = D^3 S'_{\nu'}(\nu/D)\{1-\exp[-\tau'_{\nu'}(\nu/D)]\}
\end{eqnarray}
where $\tau'_{\nu'}(\nu/D)=\alpha'_{\nu'}(\nu/D)\ell'$.

\subsection{Inverse Compton scattering of synchrotron photons (SSC)}

In the present paper I shall assume that we view the emission
emitted at angle $\theta'=\theta=0$ to the jet axis.  The
electrons are assumed to be isotropic and mono-energetic in the
jet frame, and have Lorentz factor $\gamma'=\gamma_1/D$, and
velocity $\beta'c=(1-1/{\gamma'}^2)^{1/2}c$.  For simplicity, I
shall also take the synchrotron target photons to be isotropic in the
jet frame, while using the results from Section~3 to normalize
their spectrum.

For isotropic target photons in the jet frame with frequency
$\nu'_0$, the frequency of the scattered photons range between 0
and $4 {\gamma'}^2\nu'_0$.  The jet-frame SSC emissivity (erg
cm$^{-3}$ s$^{-1}$ sr$^{-1}$ Hz$^{-1}$) is then
\begin{eqnarray}
j_{\nu'}^{\rm \prime \, SSC} = {n_e' 3 h \nu' \sigma_T c \over 4 {\gamma'}^2}
\int_{\nu'/4 {\gamma'}^2}^\infty {d \nu'_0 \over \nu'_0 }\langle n'_{\rm syn}(\nu'_0)\rangle f_{\rm IC}(x)
\end{eqnarray}
where 
\begin{eqnarray}
\langle n'_{\rm syn}(\nu_0')\rangle ={ M(\ell'/r,\tau_\nu \ll 1) \over h \nu'_0c}I_{\nu'}^{\rm \prime syn}(\nu_0',\theta'=0)
\end{eqnarray}
is the average jet-frame synchrotron specific photon number density
(photons cm$^{-3}$ sr$^{-1}$ Hz$^{-1}$), $x=\nu'/4{\gamma'}^2\nu_0'$  and $f_{\rm IC}(x) = 2x\ln x +x +1
-2x^2$ (Blumenthal and Gould 1970),
giving
\begin{eqnarray}
I_{\nu'}^{\rm \prime \, SSC}(\nu';\theta'\!=\!0) = \ell' j_{\nu'}^{\rm \prime \, SSC}.
\end{eqnarray}
Finally, this is Doppler boosted to the observer frame
\begin{eqnarray}
I_{\nu}^{\rm SSC}(\nu;\theta\!=\!0) = D^{3}I_{\nu'}^{\rm \prime \, SSC}(\nu/D; \theta'\!=\!0)
\end{eqnarray}

\subsection{Inverse-Compton scattering of the CMBR: gamma ray production}

In the jet frame, the black body temperature would depend on the
angle $\theta_0'$ with respect to the jet axis at which the CMBR
target photons propagate
\begin{eqnarray}
T'(\theta_0')={(1+z)T_0 \over \Gamma(1 + \beta_j \cos\theta_0')},
\end{eqnarray}
such that their specific photon number density
(photons cm$^{-3}$ sr$^{-1}$ Hz$^{-1}$) is
\begin{eqnarray}
n'_{\rm CMBR}(\nu'_0,\theta_0')={2 h {\nu'_0}^2/c^3 \over \exp[h {\nu'_0}/kT'(\theta_0')]-1}.
\end{eqnarray}

For target photons in the jet frame with frequency $\nu'_0$
propagating at angle $\theta_0'$ to the jet axis, the frequencies
of the photons scattered by electrons propagating parallel to the
jet axis are uniformly distributed between 0 and $\nu'_{\rm max}(\nu'_0,\theta_0')=2
{\gamma'}^2\nu'_0(1-\beta'\cos\theta_0')$ in the approximation
that the scattered photons are isotropic in the electron rest
frame.  The jet-frame emissivity (erg cm$^{-3}$ s$^{-1}$
sr$^{-1}$ Hz$^{-1}$) for IC on the CMBR in the jet direction is
then
\begin{eqnarray}
j_{\nu'}^{\rm \prime \, ICM}(\theta'\!=\!0)  &=& {n_e' h \nu' \sigma_T c }
\int_{-1}^1{d\cos\theta_0' \over 2}(1-\beta'\cos\theta_0')  \nonumber \\ & & \hspace*{2em} \int_{{\nu'}_0^{\rm min}(\theta_0')}^\infty {d \nu'_0  }{n'_{\rm CMBR}(\nu'_0,\theta_0') \over \nu'_{\rm max}(\nu'_0,\theta_0')}
\end{eqnarray}
where ${\nu'}_0^{\rm min}(\theta_0')=\nu'/2 {\gamma'}^2(1-\beta'\cos\theta_0')$.
Finally, 
\begin{eqnarray}
I_{\nu'}^{\rm \prime \, ICM}(\nu';\theta'\!=\!0) &=& \ell' j_{\nu'}^{\rm \prime \, ICM}(\theta'\!=\!0),\\
I_{\nu}^{\rm ICM}(\nu;\theta\!=\!0)  &=& D^{3}I_{\nu'}^{\rm \prime \, ICM}(\nu/D; \theta'\!=\!0).
\end{eqnarray}

\section{Discussion}

The angular radius inferred by Kedziora-Chudczer et al.\ (1997)
is $\theta_{1/2} \approx 1.5 \times 10^{-11}$~rad, i.e.\
$\log(\theta_{1/2})=-10.83$.  Examining the parameter space for
$\nu_c=10^{9.1}$ in Fig.~\ref{fig:parameters}(a), one sees that
all models with this angular radius lie well inside the region
where IC on the CMBR dominates the energy losses of electrons.
For $\eta=1$ (equal energy density in magnetic field and
relativistic electrons) and $\ell'/r=1$ this would correspond to
a point roughly mid-way between models A and B, and would require
a Doppler factor of $D \sim 5000$.  By choosing $\ell'/r > 1$ one
may reduce the Doppler factor, but unrealistically large $\ell'/r$
values are needed to reduce $D$ by a large factor.  To reduce the
Doppler factor to a reasonable value, i.e.\ $D < 100$, would
require the angular diameter to be larger by only a factor of
$\sim 5$ which, as discussed earlier, I believe is quite possible
(for an angular diameter larger by a factor of 10, even $D \sim
10$ is possible).

\subsection{Spectral energy distribution}

Because of the dominance of IC losses on the CMBR for models
A--D, one would expect the SSC peak at gamma-ray energies in the
SED to exceed the synchrotron peak for these models.  The SEDs
have been calculated for models A--E and are shown in
Fig.~\ref{fig:SED}(a) and, as expected, show strong gamma-ray
emission for models A--D.  Non-contemporaneous X-ray and
gamma-ray data are shown, and may be used with caution as upper
limits.  I have added to Fig.~\ref{fig:SED} the sensitivity of
the IBIS gamma-ray detector on INTEGRAL (Parmar et al.\ 2002), and
the expected sensitivities of GLAST (Gehrels and Michelson 1999),
and HESS (Hofmann et al.\ 2001) and CANGAROO III (Enomoto et al.\
2002), the last two being southern hemisphere atmospheric
Cherenkov telescopes nearing completion.  Taking the X-ray data
as an upper limit does not rule out any of these models.
However, the EGRET upper limit obtained from figure~3 of Hartman
et al.\ (1999) appears to exclude models A--C but be consistent
with models D--E (note, however, that the EGRET data are not
contemporaneous with the IDV data).  Models D--E have
$\theta_{1/2} \approx 6 \times 10^{-11}$--$10^{-10}$~rad and $D
\approx 180$--40, respectively.

\begin{figure}
\centerline{\psfig{file=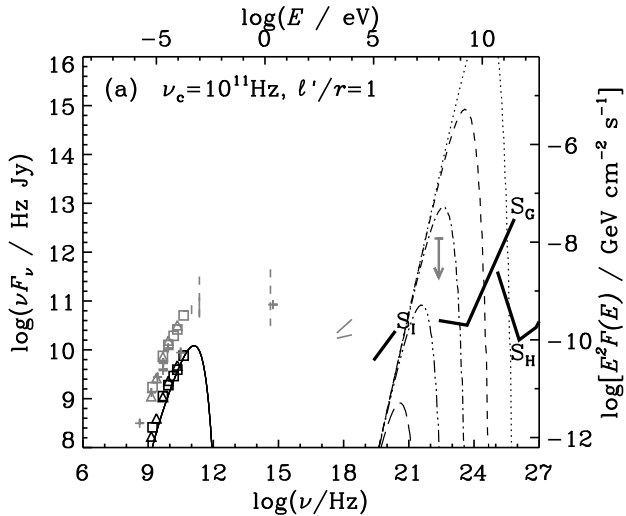, width=1.15\hsize}}
\caption{SED of IDV core of PKS~0405-385 obtained from observed
fluxes assuming flux fraction $S_c$ only for the ACTA/VLA data as
in Fig.~1.  Data additional to those shown in Fig.~1: X-ray data
at $\sim 10^{18}$~Hz -- 1997 ASCA fluxes (L.\ Kedziora-Chudczer
personal communication 1999); gamma-ray data -- EGRET upper limit
(Hartman et al.\ 1999).  Sensitivities of the IBIS instrument on
INTEGRAL (S$_{\rm I}$), GLAST (S$_{\rm G}$) and HESS (S$_{\rm
H}$) indicated (the expected sensitivity of CANGAROO III is
similar to that of HESS).  The calculated SED is shown for
various models with the synchrotron component as the solid curve
on the left, the SSC component in the middle (when present) and
the scattered CMBR component on the right.  (a) The SED for
models A--E in Fig.~\protect\ref{fig:parameters}(a) with
$\nu_c=10^{9.1}$, $\eta=1$ and $(\ell'/r)=1$: A -- dotted curves,
B -- short dashed curves, C -- chain curves, D -- triple-dot
dashed curves, E -- long dashed curves.}
\label{fig:SED}
\end{figure}

\begin{figure}
\centerline{\psfig{file=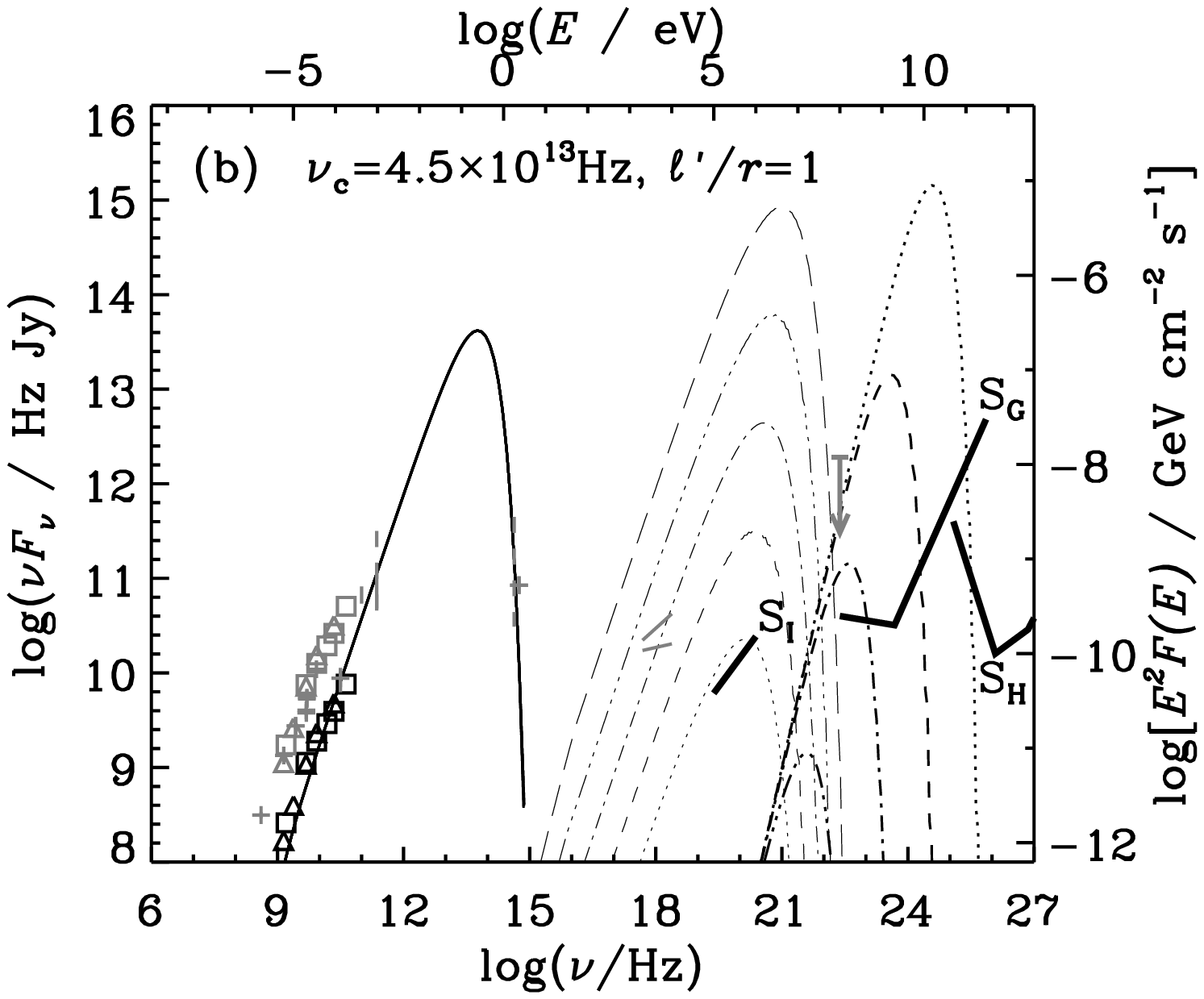, width=1.15\hsize}}
\centerline{\psfig{file=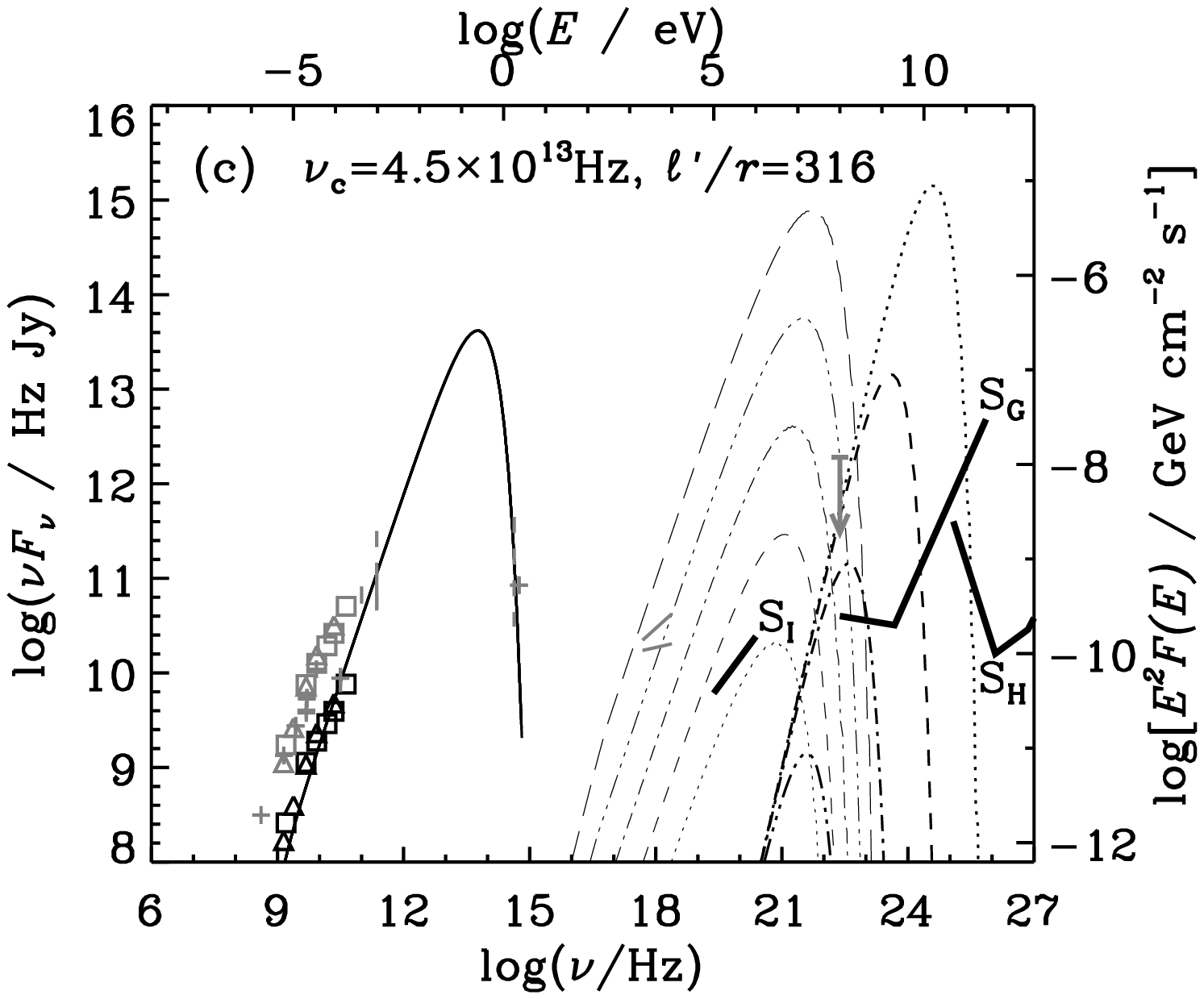, width=1.15\hsize}}
\contcaption{(b) The SED for models
F--J in Fig.~\protect\ref{fig:parameters}(b) with
$\nu_c=10^{13.65}$, $\eta=1$ and $(\ell'/r)=1$: F -- dotted
curves, G -- short dashed curves, H -- chain curves, I --
triple-dot dashed curves, J -- long dashed curves.  (c) The SED
for models K--O in Fig.~\protect\ref{fig:parameters}(b) with
$\nu_c=10^{13.65}$, $\eta=1$ and $(\ell'/r)=316$: K -- dotted
curves, L -- short dashed curves, M -- chain curves, N --
triple-dot dashed curves, O -- long dashed curves.}
\end{figure}

As expected from equation~\ref{eq:d_model}, increasing $\nu_c$
from $10^{11}$~Hz to $10^{13.65}$~Hz reduces $D$ by a factor of
10 if one keeps all other parameters fixed (note, however, that
this reduction in $D$ is accompanied by an increase in
$B'_\perp$).  This is clearly seen in
Fig.~\ref{fig:parameters}(b) which is for $\nu_c=10^{13.65}$~Hz,
and thereby presents more opportunity for obtaining reasonable
values of $D$.  For $\eta=1$ and $\ell'/r=1$, models F--J have
the same angular radius range ($10^{-11}$--$10^{-10}$~rad) as
models A--E, but with considerably lower Doppler factors, and
these models range from being in the IC on CMBR dominated regime
(model F -- expect highest gamma-ray flux, lowest SSC flux) to
being in the SSC dominated regime (model J -- expect highest SSC
flux, lowest gamma-ray flux).  The calculated SEDs for models
F--J are shown in Fig.~\ref{fig:SED}(b) and show the trends
expected.  Possibly models I and J may be ruled out by the X-ray
data taken as an upper limit (note, however, that the X-ray data
is not contemporaneous with the IDV data).  Models F--I predict
high gamma-ray fluxes from MeV energies to 100 GeV energies,
depending on model, which could be tested observationally by
INTEGRAL, GLAST, HESS and CANGAROO III.  Models F--I have
$\theta_{1/2} \approx 10^{-11}$--$6 \times 10^{-11}$~rad and $D
\approx 1000$--20, respectively.

As also expected from equation~\ref{eq:d_model}, increasing
$\ell'/r$ from 1 to 316 results in a reduction in $D$ by a factor
of $\sim 2$ (plus a reduction in $B'_\perp$) if one keeps all
other variables fixed.  Models K--O in
Fig.~\ref{fig:parameters}(b) have all other parameters the same
as models F--J, and the resulting SEDs are shown in
Fig.~\ref{fig:SED}(c).  Because of the reduction in synchrotron
photon energy density associated with $\ell'/r \gg 1$, although
not identical, the SEDs of models K--O are qualitatively similar
to those of models F--J, with model O being ruled out by both the
gamma-ray limit and the X-ray data, although again I note that
these data are not contemporaneuos with the IDV data.  Models
K--N have $\theta_{1/2} \approx 10^{-11}$--$6 \times
10^{-11}$~rad and $D \approx 400$--8.  Model M, having
$\theta_{1/2}$ only $\sim 5$ times that assumed by
Kedziora-Chudczer et al.\ (1997) has a Doppler factor of $\sim
30$, and lower Doppler factors are possible with, for example
even larger $\ell'/r$.  Bearing in mind that no attempt has been
made to fit the (non-contemporaneous) X-ray and gamma-ray data,
and that a simple mono-energetic electron spectrum has been used,
it seems that models which are able to fit the IDV radio data
tend to predict observable fluxes of X-ray, and/or gamma ray
emission at sub-GeV and/or $\sim$10 GeV energies.  Use of a more
sophisticated electron spectrum is unlikely to alter this
conclusion.  Hence, it may well be profitable for space gamma ray
telescopes such as INTEGRAL and GLAST, and southern hemisphere
atmospheric Cherenkov telescopes such as HESS and CANGAROO III to
look for emission from PKS~0405-385 and other extreme IDV radio
galaxies.

\subsection{Variability}

The rapid intra-day variability in PKS~0405-385 is almost
certainly due to interstellar scintillation effects.  The radio
flux does, however, also vary on longer timescales -- 10 per cent
change in 2 months in July 1996 (corresponding to $\sim 1$~yr for
50 per cent change), and has very intense for periods of about a
month (Kedziora-Chudczer et al.\ 1997).

A mechanism that may cause such variability would be the emission
region moving along the jet and passing a region where external
factors cause compression/expansion of the emission region,
increasing/decreasing the magnetic field, and causing adiabatic
acceleration/deceleration of relativistic charged particles,
thereby affecting the observed intensity.  Change can also occur
as a result of the emission region passing through a bend in the
jet causing, amongst other things, a change in viewing angle with
respect to the motion of the emission region, and hence a change
in Doppler factor.  

For the examples above, the observer-frame variability time
depends on the geometry of the emission region (see, e.g.,
Protheroe 2002) and, for viewing the assumed cylindrical emission
region along its axis, is at least $\ell'/(2Dc)$.  In general,
arbitrarily low Doppler factors are possible for arbitrarily high
$\ell'/r$.  However, in this case this is clearly at the expense
of having a time scale for variability which may be unreasonably
large, and of course also requires the jet to be extremely well
collimated over the length of the emission region.

Alternatively a shock may pass through the
emission region compressing the magnetic field and accelerating
particles -- a pre-existing mono-energetic population of
relativistic particles will receive an energy boost by a factor
of $\sim {\gamma'}_{\rm shock}^2$ and acquire a power-law tail as a
result of diffusive shock acceleration.  In the case of a plane shock 
moving at jet-frame speed $\beta'_{\rm shock}c$ along the jet (see, e.g. Protheroe 2002), 
the observer-frame variability time in our case ($\theta'=0$) is at least
\begin{eqnarray}
t_{\rm var, shock}=D^{-1}\left( {\ell'\over 2c}\right)\left|1-{1 \over \beta'_{\rm shock}}\right|
\end{eqnarray}
which is longer (shorter) than $\ell'/(2Dc)$ if $\beta'_{\rm
shock}$ is less than (greater than) 0.5 for a forward-moving
shock.

\begin{figure}
\centerline{\psfig{file=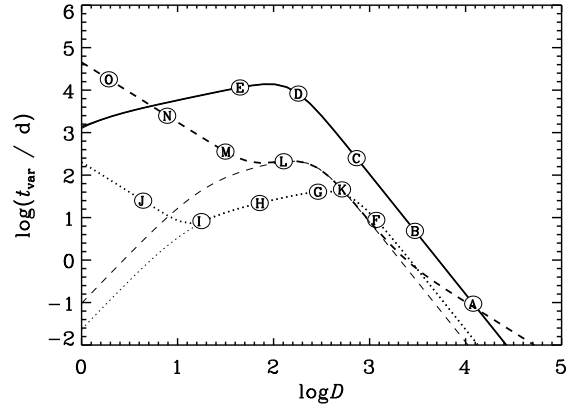, width=\hsize}}
\caption{Observer-frame variability time (thick curves) and
energy-loss time (thin curves) vs.\ Doppler factor, $D$, for:
$\nu_c=10^{9.1}$, $\eta=1$ and $(\ell'/r)=1$ (solid curves);
$\nu_c=10^{13.65}$, $\eta=1$ and $(\ell'/r)=1$ (dotted curves);
and $\nu_c=10^{13.65}$, $\eta=1$ and $(\ell'/r)=316$ (dashed
curves).  The variability times for models A--O are indicated. }
\label{fig:dopp_tvar}
\end{figure}

Except for a point-like emission region following a bent or
helical trajectory such that the Doppler boosting factor, changes
rapidly with time, the variability time can not be shorter than
the observer-frame energy-loss time scale, $D^{-1}E'/(dE'/dt')$.
This is easily obtained from equation~(\ref{eq:dedt}) and is
plotted (thin curves) against Doppler factor in
Fig.~\ref{fig:dopp_tvar} for the following cases:
$\nu_c=10^{9.1}$, $\eta=1$ and $(\ell'/r)=1$ (solid curves)
$\nu_c=10^{13.65}$, $\eta=1$ and $(\ell'/r)=1$ (dotted curves);
and $\nu_c=10^{13.65}$, $\eta=1$ and $(\ell'/r)=316$ (dashed
curves).  As noted above, the variability time will in general
be longer than the energy-loss time scale due to the dimensions
of the emission region.  For the present geometry and viewing
angle, I take the variability time to be
\begin{eqnarray}
t_{\rm var} = D^{-1}\left\{ \left( {\ell'\over
2c}\right)^2+\left[{E'\over (dE'/dt')}\right]^2\right\}^{1/2},
\end{eqnarray}
but one should note that in the case of shock excitation that it
can be longer than this for $\beta'_{\rm shock} < 0.5$, or as
short as $D^{-1}E'/(dE'/dt')$ if $\beta'_{\rm shock} \to 1$.  In
Fig.~\ref{fig:dopp_tvar}, I have also plotted $t_{\rm var}$ vs.\
$D$ for the same three cases, and the values for models A--O are
indicated by the letters.  Of the models with Doppler factors less than
$\sim 10^3$, models C--E give variability times $\sim 1$--30
years.  Models F--J give variability times of $\sim 1$ week to
$\sim $ month, and are quite compatible with the long-term
variability observed for PKS~0405-385.  Models K--O, with larger
$\ell'/r$, give variability times of $\sim 1$ month to $\sim
100$ years, with model M having $t_{\rm var} \approx 1$~year.

\section{{Conclusion}\label{sec:conclusion}}

To obtain low Doppler factors, without invoking extreme
variations from equipartition, requires a larger angular diameter
for the IDV core of PKS~0405-385 than the $\sim 6$~micro-arcsec
assumed by Kedziora-Chudczer et al.\ (1997) corresponding to a
distance of $\sim 500$~pc to the ionized material responsible for
the interstellar scintillation.  A further reduction in the
minimum Doppler factor needed to avoid the Compton Catastrophe
can be obtained by having optically thin IDV core emission,
preferably originating from a region elongated along the jet, and
observing it at a small viewing angle with respect to the jet
axis.  An angular diameter a factor of $\sim 2$--4 larger than
assumed by Kedziora-Chudczer et al.\ (1997) and Walker (1998)
corresponds roughly to models D, H--I and M--N, all of which are
allowed by the present data, and have Doppler factors $\sim 200$,
80--20 and 30--8, respectively.  Interstellar scintillation
screen distances as small as 20~pc, possibly associated with
material in the local bubble, have previously been invoked to
explain IDV in J1819+3845 (Dennett-Thorpe \& de Bruyn 2000,
2002).  By using a distance of 20~pc, even lower Doppler factors
could easily fit the observations.

In conclusion, if one observes optically thin emission along a
long narrow emission region, the average energy density in the
emission region can be significantly lower than $(4\pi/c)\int
I_\nu d\nu$.  If one takes into account the uncertainty in the
distance to the ionized clouds responsible for interstellar
scintillation causing rapid IDV in PKS~0405-385 the brightness
temperature could be as low as $\sim 10^{13}$~K, or lower, at
5~GHz.  The radio spectrum can be fit by $I_\nu \propto
\nu^{1/3}$ expected if the emission is optically thin and the
electrons are mono-energetic, or have a minimum Lorentz factor
whose critical frequency is well above the frequency range of
interest.  Such a spectrum would occur naturally in models in
which $e^\pm$ are produced as secondaries of interactions of
protons, e.g.\ by Bethe-Heitler pair production, or through
$\pi^\pm \to \mu^\pm \to e^\pm$ decay following collisions of
with matter or low-energy target photons (pion photoproduction)
by protons, or by neutrons themselves produced in pion
photoproduction interactions (e.g.\ $p\gamma \to n\pi^+$) perhaps
closer to the central engine.  The combination of all these
factors enables the Compton catastrophe to be avoided, and also
predicts that X-ray and gamma-ray emission at observable
flux levels should be expected from compact cores of extreme IDV
sources such as PKS~0405-385.

\section*{Acknowledgments}
My interest in this subject was stimulated by a workshop in
September 1999 organized by the RCfTA at the University of
Sydney.  I thank Lucyna Kedziora-Chudczer, Dave Jauncey and
Stefan Wagner for supplying data from observations of
PKS~0405-385.  This research is supported by a grant from the
Australian Research Council.

\end{document}